\documentclass[reprint,superscriptaddress,
groupedaddress,
amsmath,
amssymb,
aps,
prl,
floatfix,
]{revtex4-2}

\usepackage[pdftex]{graphicx}
\usepackage{dcolumn}
\usepackage{bm}
\usepackage[pdftex]{hyperref}
\usepackage{color}
\usepackage{here}
\usepackage{txfontsb}

\usepackage{color}
\usepackage{here}
\usepackage{txfontsb}

\begin{document}
  \title{Edge of Many-Body Quantum Chaos in Quantum Reservoir Computing}
  \author{Kaito Kobayashi}
  \author{Yukitoshi Motome}
  \affiliation{Department of Applied Physics, the University of Tokyo, Tokyo 113-8656, Japan}
  \date{\today}
  \begin{abstract}
    Reservoir computing (RC) is a machine learning paradigm that harnesses dynamical systems as computational resources. 
    In its quantum extension---quantum reservoir computing (QRC)---these principles are applied to quantum systems, whose rich dynamics broadens the landscape of information processing. 
    In classical RC, optimal performance is typically achieved at the ``edge of chaos," the boundary between order and chaos. 
    Here, we identify its quantum many-body counterpart using the QRC implemented on the celebrated Sachdev-Ye-Kitaev model. 
    Our analysis reveals substantial performance enhancements near two distinct characteristic ``edges": a temporal boundary defined by the Thouless time, beyond which system dynamics is described by random matrix theory, and a parametric boundary governing the transition from integrable to chaotic regimes. 
    These findings establish the ``edge of many-body quantum chaos" as a design guideline for QRC.
  \end{abstract}

\maketitle

{\it Introduction.}
---Quantum mechanics harbors immense potential to revolutionize modern technology~\cite{Feynman:1982}. 
Breakthroughs in quantum computing, communication, and sensing have vividly showcased the transformative power of quantum resources~\cite{Google:Nature:2019,Zhong:Science:2020,Gisin:RevModPhys:2002,Degen:RevModPhys:2017}. 
These advances have, in turn, inspired extensive research efforts in quantum machine learning, with various approaches---notably variational quantum algorithms (VQAs)---being actively explored~\cite{Peruzzo:NatComm:2014,Biamonte:Nature:2017,McClean:NatCommun:2018,Mitarai:PRA:2018,Havl:Nature:2019,Cerezo:NatRevPhys:2021}. 

More recently, the power of quantum has also integrated with reservoir computing (RC), a machine learning framework renowned for its efficacy in time-series analysis~\cite{Jaeger:GMD:2001,Jaeger:Science:2004,Verstraeten:NewralNetw:2007,Tanaka:NeuralNetw:2019,Kobayashi:SciRep:2023}. 
True to its name, the most distinctive feature of RC is the use of an input-driven dynamical system, known as a ``reservoir", to nonlinearly map input data onto a feature space. 
Notably, training in RC is limited to the post-processing stage, and the reservoir itself remains fixed, which significantly reduces the training cost. 
Extending this idea into the quantum domain, Fujii and Nakajima introduced quantum reservoir computing (QRC)~\cite{Fujii:PRAppl:2017}. 
QRC harnesses natural quantum dynamics as a reservoir, generating rich, nonlinear input-to-output mappings in an exponentially large Hilbert space without encountering the trainability bottlenecks prevalent in VQAs. 
This capability has sparked significant theoretical progress~\cite{Nakajima:PRAppl:2019,Ghosh:npjqi:2019,Pena:PRL:2021,Nokkala:CommPhys:2021,Sakurai:PRAppl:2022,Xia:FrontPhys:2022, Bravo:PRXQ:2022,Pena:CognComput:2023,Mujal:npjqi:2023,Llodra:AdvQuantTech:2023,Yasuda:arXiv:2023,Sannia:Quantum:2024,Palacios:CommPhys:2024,ifmmode:PRR:2024,Kobayashi:PRXQuantum:2024,Ivaki:PRA:2024,Llodra:Chaos:2025,Kobayashi:SciPostPhys:2025,Kobayashi:NatComm:2025,Ivaki:arXiv:2025,ivaki:arXiv:202510}, as well as promising experimental demonstrations~\cite{Negoro:arXiv:2018,Chen:PRAppl:2020,Suzuki:SciRep:2022,Kubota:PRR:2023,Hu:PRX:2023,Senanian:NatCommn:2024,Milan:arXiv:2024, Suprano:PRL:2024,Hou:arXiv:2025}. 

Given the optimization-free nature of RC, researchers naturally ask: {\it What characterizes an effective reservoir system?} 
While this question remains largely unanswered for QRC, several key criteria have been identified for classical RC. 
The most renowned among them is the concept of the ``edge of chaos"~\cite{Packard:1988,Langton:PhysicaD:1990,Bertschinger:NeuralComput:2004,Legenstein:NeuralNetw:2007,Boedecker:TheoryBiosci:2012}.  
This posits that the boundary between chaotic and nonchaotic dynamical regimes provides a performance sweet spot for RC systems, as empirically verified across diverse platforms~\cite{Takano:OptExpress:2018,Hochstetter:NatCommun:2021,Nishioka:SciAdv:2022}. 
By analogy, exploring a quantum version of the edge of chaos offers a promising route to addressing the fundamental question posed above.

In classical systems, chaos is naturally framed via phase space trajectories~\cite{Ott:2002}. 
In quantum mechanics, however, the Heisenberg uncertainty principle precludes such a direct picture, and one must instead rely on more subtle diagnostics. 
Early investigations of quantum chaos therefore focused on single-particle systems whose classical limits are chaotic, leading to the conjecture (with some analytical supports) that their spectral statistics are governed by random matrix theory (RMT)~\cite{Brody:RevModPhys:1981,BGS:PRL:1984,Berry:ProcRSocA:1985,Guhr:PhysRep:1998,Muller:PRL:2004}. 
In recent years, attention has shifted to many-body settings, spurred in large part by the exactly solvable Sachdev-Ye-Kitaev (SYK) model~\cite{Sachdev:PRL:1993,Kitaev:2015,Sachdev:PRX:2015,Fu:PRB:2016}. 
Various diagnostics, including level statistics~\cite{Oganesyan:PRB:2007,Atas:PRL:2013,Haake:Book:2018,Giraud:PRX:2022}, out-of-time-order correlators~\cite{Roberts:PRL:2015,Roberts:JHEP:2017,Maldacena:JHEP:2016}, and spectral form factors~\cite{Cotler:JHEP:2017,Cotler:JHEP:20172,Liu:PRD:2018,Gharibyan:JHEP:2018} have enriched our understanding of chaos in interacting quantum systems. 
Experimental realizations of the SYK physics have also been proposed on several platforms~\cite{Danshita:2017:PTEP,Garc:PRL:2017, Pikulin:PRX:2017, npj:Luo:2019,Jafferis:Nature:2022}.
Notably, the SYK model exhibits maximally quantum chaotic behavior, saturating the so-called bound on chaos~\cite{Polchinski:JHEP:2016,Maldacena:JHEP:2016,Maldacena:PRD:2016} and has stood as a canonical model for the investigation of many-body quantum chaos. 
These insights motivate the exploration of QRC built upon the SYK model, aiming to address the open question of reservoir effectiveness from a quantum chaotic standpoint.

In this Letter, we present a unified, RMT-based analysis that identifies the ``edge of many-body quantum chaos" as the optimal guide for operating QRC. 
In particular, we examine the QRC performance near two distinct chaotic boundaries of the SYK model. 
The first is a temporal boundary defined by the Thouless time, beyond which the system's dynamics conforms to RMT. 
The second is a parametric boundary marking the transition from quantum chaotic to integrable (nonchaotic) behavior, controlled by the strength of an additional noninteracting term. 
Using representative machine learning tasks that demand both memory retention and nonlinear transformation, we demonstrate that QRC performance is maximized near the onset of quantum chaotic regime in both temporal and parametric domains. 
These observations establish the ``edge of many-body quantum chaos" as an important guideline for the design of QRC systems.

\begin{figure}[htbp]
  \centering
  \includegraphics[width=0.85\hsize]{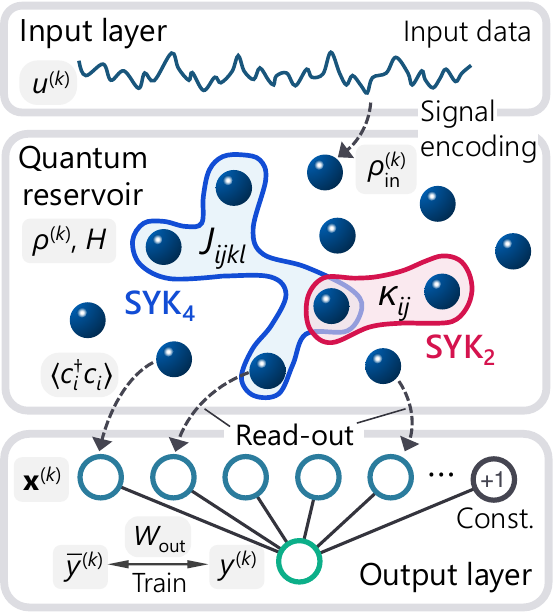}
  \caption{
    Schematic illustration of the SYK model and its implementation in QRC architecture. 
    The SYK system described in Eq.~(\ref{eq1}), which serves as the quantum reservoir, comprises the SYK\(_4\) term (\(J_{ijkl}\)) and the SYK\(_2\) term (\(\kappa_{ij}\)). 
    At each time step \(k\), the process begins by encoding an input \(u^{(k)}\) into the quantum reservoir via the input state \(\rho_{\mathrm{in}}^{(k)}\). 
    The reservoir state \(\rho^{(k)}\) then evolves under the Hamiltonian according to the update rule in Eq.~(\ref{eq2}). 
    Information is subsequently read out by constructing a state vector \(\bm{\mathrm{x}}^{(k)}\) from the expectation values \(\langle c_i^\dagger c_i\rangle\) and a constant bias term. 
    Finally, this vector is linearly transformed by the trainable weight matrix \(W_{\mathrm{out}}\) to produce the output \(y^{(k)}\), which is optimized to approximate the target \(\bar{y}^{(k)}\). 
    }
  \label{fig1}
\end{figure}

{\it Model.} 
---Figure \ref{fig1} schematically illustrates the QRC architecture implemented with the SYK model for fermions. 
The quantum reservoir system, whose dynamics is harnessed for information processing, is governed by the Hamiltonian 
\begin{equation}
\mathcal{H}=\sum_{i,j,k,l=1}^NJ_{ijkl}c^\dagger_ic^\dagger_j c_k c_l + \sum_{i,j=1}^N \kappa_{ij}c^\dagger_i c_j, \label{eq1}
\end{equation}
where \(N\) denotes the system size, and \(c^\dagger_i\) (\(c_i\)) is the spinless fermionic creation (annihilation) operator at site \(i\). 
In Eq.~(\ref{eq1}), the first term describes the SYK\(_4\) interaction: the all-to-all couplings \(J_{ijkl}\) are drawn from a zero-mean complex Gaussian distribution of variance \(\overline{|J_{ijkl}|^2}=J_4^2/N^3\) (\(\overline{\bullet}\) denotes the ensemble average).
The second term is the noninteracting SYK\(_2\) contribution, with coupling constants \(\kappa_{ij}\) sampled from a zero-mean complex Gaussian distribution of variance \(\overline{|\kappa_{ij}|^2}=\kappa_2^2/(2N)\). 
The couplings satisfy the symmetry relations \(\kappa_{ij}=\kappa_{ji}^*\), \(J_{ijkl}=J_{klij}^*\), and \(J_{ijkl}=-J_{jikl}=-J_{ijlk}\). 
Both coupling terms conserve the total fermion number \(N_p\equiv \sum_i \langle c_i^\dagger c_i \rangle\). 
Unless otherwise specified, we set \(N=8\), \(J_4=1\), and \(N_p=N/2\) (note that the QRC input protocol introduced below does not conserve \(N_p\)).

\begin{figure}[b]
  \centering
  \includegraphics[width=\hsize]{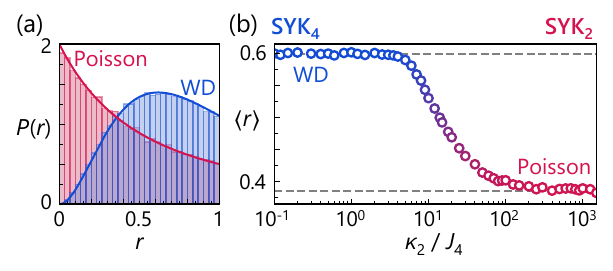}
  \caption{
    (a) Distribution of level spacing ratio \(P(r)\) for the SYK\(_4\) model (blue) and the SYK\(_2\) model (red). 
    Corresponding WD and Poisson predictions are shown in blue and red curves, respectively. 
    (b) Level spacing ratio \(\langle r \rangle\) as a function of the coupling strength \(\kappa_2/J_4\), computed over the central \(50\%\) of the spectrum. 
    The horizontal dashed lines represent the reference values for Poisson and WD statistics. 
    In both panels, data are averaged over \(2000\) realizations for the system size \(N=8\). 
    }
  \label{fig2}
\end{figure}

To diagnose the chaotic nature of quantum many-body systems, we first examine the level statistics~\cite{Oganesyan:PRB:2007,Atas:PRL:2013,Haake:Book:2018,Giraud:PRX:2022}. 
Let the ordered eigenvalues of \(\mathcal{H}\) be \(E_{n}\); the spacings between consecutive energy levels are given by \(s_n = E_{n+1} - E_{n}\). 
According to RMT, the spacing distribution in integrable systems follows Poisson statistics, whereas in quantum chaotic systems, it is well described by Wigner-Dyson (WD) distribution. 
A convenient metric to characterize these distributions is the level spacing ratio, defined as \(r_n=\min(s_n/s_{n+1}, s_{n+1}/s_{n})\). 
Its mean value \(\langle r\rangle\) assumes \(\approx 0.3862\) for Poisson distribution and \(\approx 0.5996\) for WD statistics~\cite{Oganesyan:PRB:2007,Atas:PRL:2013}. 
Figure \ref{fig2}(a) presents the distribution of the level spacing ratio \(P(r)\) for the SYK\(_4\) model [\((J_4,\kappa_2)=(1,0)\)] and the SYK\(_2\) model [\((J_4,\kappa_2)=(0,1)\)], in excellent agreement with the WD and Poisson predictions, respectively. 
Notably, when both terms are present, the system can be tuned between these extremes: as the ratio of the typical coupling strengths \(\kappa_2/J_4\) increases, Fig.~\ref{fig2}(b) displays a gradual shift of \(\langle r\rangle\) from the WD to the Poisson value. 
This result strongly suggests a transition from quantum-chaotic to integrable behavior~\cite{Garcia:PRL:2018,Nosaka:JHEP:2018,Sorokhaibam:JHEP:2020,Sun:PRB:2021,Monteiro:PRR:2021}; for finite-size effects, see the Supplemental Material~\cite{suppl}.

{\it Framework of the QRC.}
---We now describe the QRC framework for information processing; all procedures employed herein follow the conventional protocols established in earlier investigations~\cite{Fujii:PRAppl:2017,Nakajima:PRAppl:2019}. 
The QRC architecture comprises four principal stages: signal encoding, reservoir dynamics, read-out, and post-processing, through which the input at the \(k\)-th step \(u^{(k)}\) is processed in an attempt to produce its corresponding target value \(\overline{y}^{(k)}\) (Fig.~\ref{fig1}). 
Consider a discretized input sequence \(\{u^{(k)}\}\equiv\{u^{(0)}, u^{(1)},\dots\}\) with \(u^{(k)}\in [0,1]\). 
Each input \(u^{(k)}\) is encoded into the reservoir by replacing the quantum state at site \(1\) with \(\rho_{\mathrm{in}}^{(k)}\equiv|\psi_{\mathrm{in}}^{(k)}\rangle\langle\psi_{\mathrm{in}}^{(k)}|\), where \(|\psi_{\mathrm{in}}^{(k)}\rangle\equiv\sqrt{1-u^{(k)}}|0\rangle + \sqrt{u^{(k)}}|1\rangle\). 
Subsequently, the system evolves unitarily under the Hamiltonian in Eq.~\eqref{eq1}. 
The encoding operation is repeated at intervals \(\Delta t_{\mathrm{in}}\), which determines the intrinsic time-scale of the QRC system.  
The overall update rule is summarized by the completely positive trace-preserving map: 
\begin{equation}
    \rho^{(k)}(t) = e^{-i\mathcal{H}t} \left[\rho_{\mathrm{in}}^{(k)} \otimes \mathrm{Tr}_1\{\rho^{(k-1)}(\Delta t_{\mathrm{in}})\}\right] e^{ i\mathcal{H}t}, \label{eq2}
\end{equation}
where \(\mathrm{Tr}_1\{\bullet\}\) denotes the partial trace over site \(1\); see the Supplemental Material for details on the observed dynamics~\cite{suppl}. 
During the evolution, the expectation value of the single-site particle number is measured at discrete times \(t_v=v\Delta t_{\mathrm{in}}/V\), where \(v=1,\dots ,V\) and \(V\) is the number of virtual computational nodes (here, \(V=10\)); for the effect of statistical noise, see the Supplemental Material~\cite{suppl}. 
At the \(k\)-th step, the outcomes \(\mathrm{Tr}\{\rho^{(k)}(t_v)c_i^\dagger c_i\}\) are assembled into a one-dimensional state vector \(\bm{\mathrm{x}}^{(k)}\); including an additional constant term, \(\bm{\mathrm{x}}^{(k)}\) has \(NV+1\) components in total. 
The output layer then applies a linear transformation \(y^{(k)}=W_{\mathrm{out}}\bm{\mathrm{x}}^{(k)}\), where the output weight \(W_{\mathrm{out}}\) is trained by minimizing the mean squared error between the computed output \(y^{(k)}\) and the target output \(\bar{y}^{(k)}\). 
In our simulations, a total of \(10000\) inputs are processed. 
Starting from a random mixed state, the first \(4000\) inputs are discarded to wash out the effects of initial conditions (see the Supplemental Material for the convergence property~\cite{suppl}). 
The subsequent \(3000\) inputs are employed for training \(W_{\mathrm{out}}\), and the final \(3000\) are reserved for evaluation. 
Performance results are averaged over \(500\) independent random realizations.

Benchmark tasks in reservoir computing are typically characterized by their requirements on temporal memory and nonlinear processing: the former specifies how far back in the input history the reservoir must retain information, while the latter is often described by the maximum polynomial order and the richness of cross-term interactions. 
Accordingly, we evaluate the reservoir performance on two complementary tasks: the short-term memory (STM) task, which probes linear memory capacity, and the nonlinear auto-regressive moving average (NARMA) task, which evaluates combined memory and nonlinear processing capabilities~\cite{Verstraeten:NewralNetw:2007,Fujii:PRAppl:2017}. 
Since real-world applications hinge on both properties, we place greater emphasis on the NARMA performance; for additional tasks, see the Supplemental Material~\cite{suppl}.

In the STM task, the target output at delay \(d\) is defined as the input value \(d\) steps earlier in a random input sequence \(\{u^{(k)}\}\), i.e., \(\bar{y}^{(k)}_d = u^{(k-d)}\). 
During the testing phase, we collect the outputs \(\bm{\mathrm{y}}=\{y^{(k)}\}\) and targets \(\bar{\bm{\mathrm{y}}}_d=\{\bar{y}^{(k)}_d\}\), and compute the determination coefficient \(R^2_d\equiv \mathrm{cov}^2(\bm{\mathrm{y}},\bar{\bm{\mathrm{y}}}_d)/[\sigma^2(\bm{\mathrm{y}})\sigma^2(\bar{\bm{\mathrm{y}}}_d)]\), where \(\mathrm{cov}\) and \(\sigma^2\) denote the covariance and the variance, respectively. 
\(R^2_d\) is bounded between \(0\) and \(1\); values closer to \(1\) indicate better memory performance. 

In the NARMA task, the target sequence at a given order \(n\) is defined by the recursive relation: \(\bar{y}^{(k+1)} = \alpha \bar{y}^{(k)} + \beta \bar{y}^{(k)}\sum_{j=0}^{n-1}\bar{y}^{(k-j)} + \gamma u^{(k-n+1)}u^{(k)} + \delta\), where (\(\alpha\), \(\beta\), \(\gamma\), \(\delta\)) are set to (\(0.3\), \(0.05\), \(1.5\), \(0.1\))~\cite{Fujii:PRAppl:2017,Pena:PRL:2021}. 
The input stream is sampled from the range \([0,0.2]\) to prevent divergence in \(\{\bar{y}^{(k)}\}\), which is then rescaled to \([0,1]\) when encoded to the quantum reservoir~\cite{Fujii:PRAppl:2017}. 
Performance is evaluated in terms of the deviation between the output and the target, which is quantified by the normalized mean-squared error (NMSE) \(\|\bm{\mathrm{y}}-\bar{\bm{\mathrm{y}}} \|^2 / \|\bar{\bm{\mathrm{y}}} \|^2\) with \(\|\bullet\|\) being the Euclidean norm. 
A lower NMSE indicates superior performance in accurately reproducing the target nonlinear dynamics.

\begin{figure}[t]
  \centering
  \includegraphics[width=\hsize]{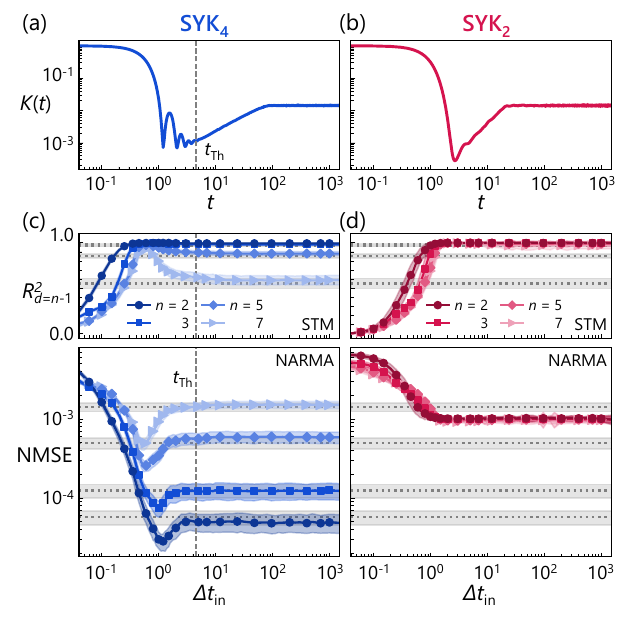}
  \caption{
    (a), (b) The SFF \(K(t)\) for the SYK\(_4\) and SYK\(_2\) models, with each curve averaged over \(20000\) realizations. 
    (c), (d) The QRC performances as functions of \(\Delta t_{\mathrm{in}}\) for the SYK\(_4\) and SYK\(_2\) models. 
    The upper panels show the memory performance \(R^2_{d=n-1}\) (higher values indicate better retention), while the lower panels display the NMSE on the NARMA tasks (lower values indicate better accuracy). 
    Markers represent the orders \(n=2,3,5\), and \(7\) (plotted sparsely for brevity). 
    The horizontal dotted lines indicate the reference performance of the Haar QRC model, with poorer performance corresponding to higher \(n\). 
    Performance results are averaged over \(500\) realizations, with shaded regions denoting the standard deviation. 
    Vertical dashed lines in (a) and (c) represent the Thouless time \(t_{\mathrm{Th}}\).
    }
  \label{fig3}
\end{figure}

{\it Temporal edge of many-body quantum chaos.}
---We begin with an examination of the QRC performance of the SYK\(_4\) and SYK\(_2\) models, in which clear fingerprints of many-body quantum chaos manifest themselves in the temporal domain. 
As a time-domain probe of the energy-spectrum correlations characteristic of quantum chaos, we analyze the spectral form factor (SFF) \(K(t) = \overline{\sum_{m,n}e^{i(E_m-E_n)t}}/\mathcal{N}^2\), where \(\mathcal{N}\) denotes the Hilbert space dimension~\cite{Cotler:JHEP:2017,Cotler:JHEP:20172,Liu:PRD:2018,Gharibyan:JHEP:2018}. 
In many-body chaotic systems, the SFF generally exhibits a ``ramp-plateau" profile governed by two time-scales: the Thouless time \(t_{\mathrm{Th}}\), after which the evolution follows RMT predictions, leading to a universal ``ramp" reflecting level repulsion, and the plateau time \(t_{\mathrm{p}}\), beyond which \(K(t)\) saturates at a plateau due to the discreteness of the spectrum~\cite{Bertini:PRL:2018,Chan:PRL:2018,Dong:PRL:2025}. 
Figures \ref{fig3}(a) and \ref{fig3}(b) illustrate the SFF for the SYK\(_4\) and SYK\(_2\) limits, respectively. 
In the SYK\(_4\) case, the SFF exhibits the pronounced ramp beginning around \(t_{\mathrm{Th}}\sim 4.5\) and the subsequent plateau beyond \(t_{\mathrm{p}}\sim 80\), providing a clear signature of many-body quantum chaos. 
While \(K(t)\) in the SYK\(_2\) model also features a dip and a plateau, they occur on a much shorter timescale and with a different system-size scaling of \(t_{\mathrm{p}}\) compared to the chaotic system (see the Supplemental Material for details~\cite{suppl}).

Figures \ref{fig3}(c) and \ref{fig3}(d) plot the QRC performance of the SYK\(_4\) and SYK\(_2\) models as functions of the input interval \(\Delta t_{\mathrm{in}}\), which defines the intrinsic time-scale governing the QRC dynamics. 
The upper panels display the STM performance \(R^2_{d=n-1}\) at delay \(d=n-1\), and the lower panels show the NMSE on the NARMA-\(n\) tasks of orders \(n=2\), \(3\), \(5\), and \(7\). 
The chosen STM delay corresponds to the oldest input term explicitly appearing in the definition of the NARMA-\(n\) dynamics; additional past inputs are nevertheless recursively encoded within the target sequence \(\{\bar{y}^{k}\}\) (see the Supplemental Material for the full delay dependence of \(R^2_d\)~\cite{suppl}). 
To establish a baseline for comparison, we also analyze the Haar QRC model, in which the evolution operator \(e^{-i\mathcal{H}t}\) is replaced by a unitary \(U\) drawn from the Haar measure; no intrinsic time scale exists therein~\cite{Innocenti:CommunPhys:2023,Vetrano:npjqi:2025}. 
Its performance, shown as horizontal dashed lines, deteriorates monotonically with increasing \(n\) as the task complexity grows.

A striking pattern emerges in the QRC performance for the SYK\(_4\) model [Fig.~\ref{fig3}(c)]. 
When \(\Delta t_{\mathrm{in}}\) is very small, the time evolution operator of the quantum reservoir is nearly the identity, failing to enrich the input-to-output map and resulting in poor performance on both the STM and NARMA tasks. 
However, as \(\Delta t_{\mathrm{in}}\) increases toward the Thouless time \(t_{\mathrm{Th}}\), the QRC performance improves dramatically and reaches its maximum just before \(t_{\mathrm{Th}}\). 
Beyond this point, both metrics level off at the baseline of the Haar QRC model for each order \(n\), in accordance with the onset of universal RMT descriptions typical of quantum chaotic systems. 
In other words, the Thouless time effectively marks the temporal boundary where fully quantum chaotic signatures begin to dominate the QRC performance. 
The observed dip of the NMSE, situated just prior to \(t_{\mathrm{Th}}\), thus explicitly underscores the performance enhancement at the ``edge of many-body quantum chaos" in the temporal domain.

\begin{figure}[b]
  \centering
  \includegraphics[width=\hsize]{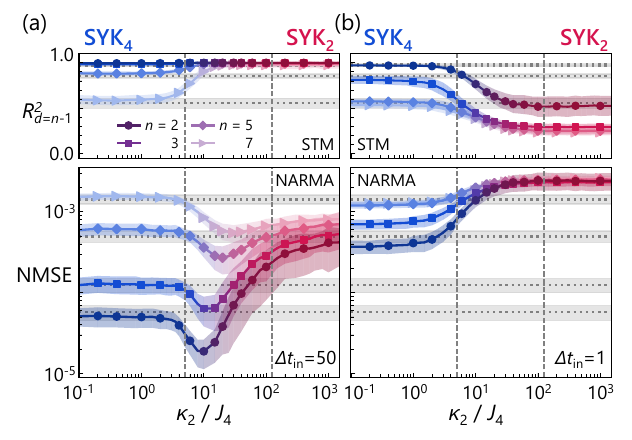}
  \caption{
    (a), (b) The QRC performance as a function of \(\kappa_2 / J_4\) for the input interval \(\Delta t_{\mathrm{in}}=50\) and \(\Delta t_{\mathrm{in}}=1\), respectively, analogous to Figs.~\ref{fig3}(c) and \ref{fig3}(d). 
    Sparsely placed markers denote the order \(n\), with colors indicating \(\langle r \rangle\) according to the color scale in Fig.~\ref{fig2}(b).
    The vertical dashed lines indicate the approximate boundaries where the absolute difference between the measured \(\langle r\rangle\) and its theoretical value for WD (left) and Poisson (right) statistics first falls below \(10^{-2}\). 
    }
  \label{fig4}
\end{figure}

In contrast, the QRC with the integrable SYK\(_2\) model behaves quite differently [Fig.~\ref{fig3}(d)]. 
As \(\Delta t_{\mathrm{in}}\) increases, the NARMA performance improves only modestly from its initial low values and quickly saturates, showing no pronounced peak. 
This behavior starkly contrasts with that of the SYK\(_4\) model, indicating that the performance enhancements observed in Fig.~\ref{fig3}(c) are fundamentally attributable to the latter's intrinsic many-body quantum chaotic nature. 
Notably, despite its poor NARMA performance, the SYK\(_2\) model attains higher STM performance than the SYK\(_4\) model. 
These model-dependent behaviors motivate our exploration of interpolations between the SYK\(_2\) and SYK\(_4\) models.

{\it Parametric edge of many-body quantum chaos.}
---We here demonstrate the QRC performance across the quantum chaotic-integrable transition by mixing the SYK\(_4\) and SYK\(_2\) Hamiltonians. 
To ensure a fair comparison as the coupling strength is varied, each Hamiltonian is rescaled by its spectral norm: \(\mathcal{H}\rightarrow\mathcal{H}/||\mathcal{H}||\)~\cite{Rossini:PRL:2020}. 
(Note that this correspondingly changes the time-scale relative to Fig.~\ref{fig3}.) 
Figures \ref{fig4}(a) and \ref{fig4}(b) display the STM and NARMA performances versus the ratio \(\kappa_2/J_4\) for long (\(\Delta t_{\mathrm{in}} =50\)) and short (\(\Delta t_{\mathrm{in}}=1\)) input intervals, respectively. 
The vertical dashed lines serve as a guide of the eye for the onset of WD and Poisson statistics, while the horizontal dotted lines represent the performance of the Haar QRC model for reference. 
We note that these input intervals are chosen to probe two distinct time-scales: the longer input interval \(\Delta t_{\mathrm{in}} =50\) evaluates QRC performance within the universal RMT regime, while the shorter interval \(\Delta t_{\mathrm{in}} =1\) characterizes QRC behavior in the nonuniversal, short-time window outside the scope of RMT. 

The most pronounced trend is observed for \(\Delta t_{\mathrm{in}} = 50\) [Fig.~\ref{fig4}(a)]. 
When \(\kappa_2/J_4\) is varied, the NARMA performance exhibits a clear dip in the NMSE at a coupling strength just before the onset of the many-body quantum chaotic regime. 
This dip provides another manifestation of the ``edge of many-body quantum chaos", in this case in the parameter domain. 
By contrast, for the short input interval \(\Delta t_{\mathrm{in}} = 1\), varying the coupling strength yields no performance enhancement [Fig.~\ref{fig4}(b)]. 
At such a short time-scale, the QRC does not reflect chaotic characteristics, so the performance changes only gradually, without any distinct peak at the chaos boundary. 

{\it Discussions.} 
---
In the strongly chaotic regime with large \(\Delta t_{\mathrm{in}}\), the QRC provides a highly nonlinear mapping, but the intrinsic featurelessness of the quantum chaotic dynamics makes input-specific signatures relatively indistinct. 
Conversely, in the nonchaotic regime or with moderately small \(\Delta t_{\mathrm{in}}\), the QRC retains input memory but sacrifices nonlinear expressivity. 
At the edge of many-body quantum chaos, the expressivity of the input-to-output map is optimally balanced against the preservation of input-specific features, leading to enhanced performance. 
From this picture, it follows that the precise location of this edge and the achievable gain depend on the task. 
Tasks that demand long memory shift the performance peak toward the nonchaotic side (e.g., the STM task with large \(d\)), whereas tasks that require stronger nonlinearity at fixed memory shift it toward the chaotic side (e.g., the nonlinear STM task). 
In extreme cases---when a task overwhelmingly requires either memory or nonlinearity, or when its complexity exceeds the QRC's capacity---the edge-induced performance enhancement may not appear. 
See the Supplemental Material for additional performance evaluations~\cite{suppl}.

Recent advances in QRC have reported related performance enhancements, and we believe that a concise comparison highlights the significance of our contributions. 
One of the authors and his collaborators identified a performance peak near the operational regime boundary in their feedback-driven QRC protocol~\cite{Kobayashi:PRXQuantum:2024}; however, this enhancement appears predominantly due to classical instabilities in feedback connections, rather than intrinsically quantum phenomena. 
In the purely quantum regime, Martínez-Peña {\it et al.} demonstrated enhanced performance near the ergodic-localization transition in the spin-based QRC~\cite{Pena:PRL:2021,Xia:FrontPhys:2022, Ivaki:PRA:2024}. 
While this finding highlights the importance of dynamical phases in QRC, the role of many-body quantum chaos, and its relation to RMT, remain largely unexplored. 
Recently, Llodrà {\it et al.} analyzed QRC performance in connection with quantum chaos appearing at the quantum phase transition in a clean system, which is complementary to our study on random systems~\cite{Llodra:Chaos:2025}. 
From a standpoint of temporal domain, investigations into quantum chaos remain sparse: Vetrano {\it et al.} explored the role of scrambling times, though within a different framework (quantum extreme learning machines)~\cite{Vetrano:npjqi:2025}.

Building upon these studies, our work directly demonstrates the relationship between quantum chaos and QRC. 
Using the SYK model---a canonical many-body quantum chaotic system---as a testbed, we have extended the concept of the ``edge of chaos" into the quantum many-body realm by uncovering two distinct edges: one is the ``temporal edge", which emerges before the Thouless time, and the other is the ``parametric edge", which appears near the quantum chaotic-to-integrable transition. 
Since our RMT-based analysis makes no system-specific assumptions, we expect the observed ``edge of many-body quantum chaos" to be broadly observable across platforms. 
Our study therefore provides clear design guidelines for engineering quantum reservoirs, unlocking advanced information processing with quantum many-body systems.

{\it Acknowledgments.}
---The authors thank Shusei Wadashima for fruitful discussions. 
This work was supported by the JSPS KAKENHI (No.~JP25H01247). 
K.K. was supported by the Program for Leading Graduate Schools (MERIT-WINGS) and JST BOOST (No.~JPMJBS2418). 
Part of computation in this work has been done using the facilities of the Supercomputer Center, the Institute for Solid State Physics, the University of Tokyo. 

\bibliography{bibtex}

\clearpage
\setcounter{equation}{0}
\setcounter{figure}{0}
\setcounter{table}{0}
\renewcommand{\theequation}{S\arabic{equation}}
\renewcommand{\thefigure}{S\arabic{figure}}
\setcounter{section}{0}
\setcounter{secnumdepth}{1}
\setcounter{secnumdepth}{2}

\widetext
\begin{center}
{\large Supplemental Material for}

\textbf{\large Edge of Many-Body Quantum Chaos in Quantum Reservoir Computing}

\vskip\baselineskip
Kaito Kobayashi\(^*\) and Yukitoshi Motome
\par
{\it Department of Applied Physics, the University of Tokyo, Tokyo 113-8656, Japan} 
\par
(Dated: \today)
\par
\(^*\)Corresponding author. E-mail: kaito-kobayashi92@g.ecc.u-tokyo.ac.jp
\end{center}
\setcounter{page}{1}

\twocolumngrid
\begin{NoHyper}

\section{Detailed characterization of many-body quantum chaos}

In the main text, we characterize many-body quantum chaos by analyzing both the energy-level statistics and the spectral form factor (SFF). 
Here, we supplement those results with additional numerical data to provide a more comprehensive discussion.

\subsection{Level statistics}

The statistics of adjacent energy-level spacings provide a clear diagnostic of quantum chaos within the framework of random matrix theory (RMT). 
In particular, an integrable system exhibits a Poissonian distribution, whereas a quantum-chaotic system follows Wigner–Dyson (WD) statistics. 
The WD class splits further into three classes according to the fundamental symmetries of the Hamiltonian: Gaussian orthogonal ensemble (GOE), Gaussian unitary ensemble (GUE), and Gaussian symplectic ensemble (GSE), each labeled by the Dyson index \(\beta=1\), \(2\), and \(4\). 
Their distributions of the level spacing ratio \(P(r)\) are given by 
\begin{equation}
    P_{\mathrm{Poi}}(r) = \frac{2}{(1+r)^2}, \quad P^{\beta}_{\mathrm{WD}}(r) = \frac{2}{Z_\beta}\frac{(r+r^2)^\beta}{(1+r+r^2)^{1+3\beta/2}},
\end{equation}
with normalization constants \(Z_1 = 8/27\), \(Z_2 = 4\pi/(81\sqrt{3})\), and \(Z_4 = 4\pi/(72\sqrt{3})\)~\cite{Atas:PRL:2013}. 
The corresponding mean values are \(\langle r\rangle_{\mathrm{Poi}}\) = \(0.3862\), \(\langle r\rangle_{\mathrm{GOE}}\) = \(0.5307\), \(\langle r\rangle_{\mathrm{GUE}}\) =\(0.5996\), and \(\langle r\rangle_{\mathrm{GSE}}\) = \(0.6744\). 

In the SYK\(_4\) limit for the Hamiltonian in Eq.~(1) of the main text, the level statistics adhere to GUE, as demonstrated by the agreement between the GUE prediction and the numerical data in Fig.~2(a) of the main text. 
This symmetry class arises because, for any finite \(N\), particle-hole symmetry (PHS) is generally broken in the employed Hamiltonian. 
To see the effect of the symmetry, we restore PHS by incorporating the correction term 
\begin{equation}
    \mathcal{H}^{\mathrm{PHS}}=\frac{1}{2}\sum_{i,j,k,l=1}^N J_{ijkl} (\delta_{i,k}c^\dagger_j c_l - \delta_{i,l}c^\dagger_j c_k- \delta_{j,k}c^\dagger_i c_l + \delta_{j,l}c^\dagger_i c_k),
\end{equation}
where \(\delta_{i,j}\) is the Kronecker delta function~\cite{Fu:PRB:2016,Rossini:PRL:2020}. 
Figure S1(a) displays the distribution \(P(r)\) in the half-filling sector \(N_p=N/2\) for the SYK\(_4\) model with the correction term, i.e., \(\mathcal{H}+\mathcal{H}^{\mathrm{PHS}}\) at \((J_4,\kappa_2)=(1,0)\). 
It now follows GOE statistics, different from the GUE behavior in Fig.~2(a) of the main text, reflecting the recovery of PHS. 
Naturally, \(P(r)\) for the SYK\(_2\) model follows Poisson statistics regardless of the PHS correction term. 
We also present the distribution for a system size \(N=7\) in Fig.~S1(b), which reintroduces GUE behavior since PHS cannot be preserved for odd system sizes. 
Likewise, the inclusion of the SYK\(_2\) term breaks PHS; even before the system reaches the integrable regime with Poisson distribution, \(P(r)\) exhibits a gradual shift from GUE to GOE as the coupling strength \(\kappa_2/J_4\) increases~\cite{Sun:PRB:2021}. 
Since our primary aim is to explore the impact of many-body quantum chaos on information processing, symmetry-related discussion specific to the SYK models is out of the current scope. 
We thus utilized the Hamiltonian in Eq.~(1) of the main text, which breaks PHS by construction. 

\begin{figure}[b]
  \centering
  \includegraphics[width=0.85\hsize]{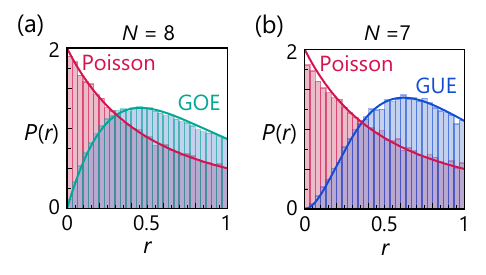}
  \caption{
    (a), (b) Distribution of level spacing ratio \(P(r)\) for the Hamiltonian including the PHS preserving term: \(\mathcal{H} + \mathcal{H}^{\mathrm{PHS}}\). 
    Panel (a) corresponds to a system size of \(N=8\), while panel (b) uses \(N=7\). 
    In both cases, the particle number sector is fixed at \(N_p = \lfloor N/2\rfloor\). 
    Theoretical predictions for Poisson (red), GOE (green), and GUE (blue) statistics are shown as curves.
    The histograms present \(P(r)\) for the SYK\(_2\) model (red) and the SYK\(_4\) model [green for (a) and blue for (b)]. 
    Data are averaged over \(2000\) independent realizations in both panels. 
    }
  \label{figS1}
\end{figure}

As illustrated in Fig.~2(b) of the main text, continuously interpolating between the SYK\(_4\) and SYK\(_2\) models drives the system from a many-body quantum chaotic phase into an integrable regime. 
To elucidate this behavior in greater detail, Fig.~S2(a) presents the full distribution \(P(r)\) for several intermediate values of the coupling ratio \(\kappa_2 / J_4\), extending the results presented in Fig.~2(a) of the main text. 
As \(\kappa_2 / J_4\) increases, \(P(r)\) evolves smoothly from the WD (GUE) statistics toward the Poisson distribution. 
In Fig.~S2(b), we plot the mean ratio \(\langle r \rangle\) as a function of \(\kappa_2/J_4\) for various system sizes at \(N_p = \lfloor N/2 \rfloor\). 
The chaotic-integrable transition is consistently observed, although a precise determination of the boundary would require even larger system sizes~\cite{Garcia:PRL:2018}. 
For the purposes of this work, we provisionally define the boundary of the quantum chaotic regime by the coupling strength at which \(\langle r\rangle\) first deviates from its RMT value. 
This reference point is used to discuss the parametric edge of many-body quantum chaos in Figs.~4(a) and 4(b) of the main text.

\begin{figure}[t]
  \centering
  \includegraphics[width=\hsize]{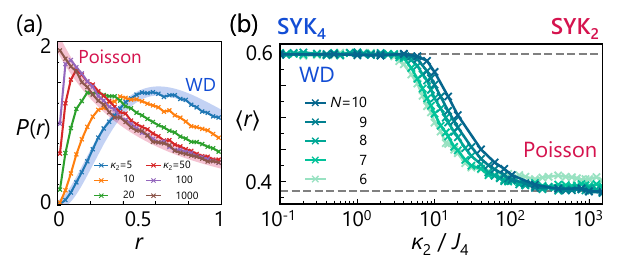}
  \caption{
    (a) Distribution \(P(r)\) for the Hamiltonian in Eq.~(1) of the main text, with a fixed system size of \(N=8\).     
    Curves are shown for several values of the coupling \(\kappa_2\) at \(J_4=1\). 
    The background blue and red curves indicate WD (GUE) and Poisson predictions. 
    Data are averaged over \(2000\) realizations.
    (b) Level spacing ratio \(\langle r\rangle\) as a function of the coupling strength \(\kappa_2/J_4\) for various system sizes. 
    The particle number sector is fixed to \(N_p = \lfloor N/2 \rfloor\). 
    The dashed gray lines mark the reference value for WD (GUE) and Poisson statistics. 
    Averages are taken over \(2000\) realizations for \(N=6,7,8\), \(1000\) for \(N=9\), and \(500\) for \(N=10\).
    }
  \label{figS2}
\end{figure}

\subsection{Spectral form factor}

\begin{figure}[b]
  \centering
  \includegraphics[width=\hsize]{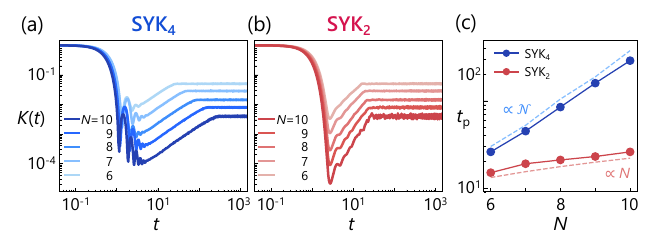}
  \caption{
    (a), (b) The SFF for the SYK\(_4\) model and the SYK\(_2\) model, shown for various system sizes. 
    Data are averaged over \(20000\) realizations for \(N=6,7,8\), \(10000\) for \(N=9\), and \(2500\) for \(N=10\).
    (c) The plateau time \(t_{\mathrm{p}}\) of the SFF as a function of the system size \(N\). 
    The blue and red dashed lines indicate reference scaling behaviors, \(\mathcal{N}=\binom{N}{N_p}\) and \(N\), respectively. 
    }
  \label{figS3}
\end{figure}

The SFF \(K(t)\) is a temporal domain probe used to capture the energy spectrum correlation across different energy scales. 
It is defined as the Fourier transform of the two level correlation function:
\begin{equation}
    K(t) =  \frac{1}{\mathcal{N}^2}\overline{\sum_{m,n}e^{i(E_m-E_n)t}},\label{Seq3} 
\end{equation}
where the sum runs over the eigenvalues with the particle number \(N_p\), and \(\mathcal N = \binom{N}{N_p}\) denotes the dimension of the corresponding Hilbert space. 
As discussed in the main text, the SFF exhibits universal characteristics in the quantum chaotic systems: at early times, it decreases due to the phase interference, while at sufficiently late times, it shows a linear ramp before the saturation to a plateau value~\cite{Bertini:PRL:2018,Chan:PRL:2018,Dong:PRL:2025}. 
In Figs.~\ref{figS3}(a) and \ref{figS3}(b), we present the SFF for the SYK\(_4\) and SYK\(_2\) models, respectively, for various system sizes. 
The ramp-plateau behavior is clearly demonstrated especially for the former system, confirming its many-body quantum chaotic nature. 
In contrast, the SFF for the latter model shows the dip and plateau over a much narrower time window. 
Indeed, Fig.~S3(c) shows that the plateau time \(t_{\mathrm{p}}\) scales with the system size as \(t_{\mathrm{p}}\sim\mathcal{N}\) for the SYK\(_4\) system and as \(t_{\mathrm{p}}\sim N\) for the SYK\(_2\) system~\cite{Cotler:JHEP:2017,Dong:PRL:2025}. 
This contrast further highlights the difference between the quantum chaotic and integrable models. 
We note that the Thouless time in the SYK\(_4\) model is known to scale as \(t_{\mathrm{Th}}\sim\sqrt{\mathcal{N}}\)~\cite{Cotler:JHEP:2017}, but the available sizes here are insufficient to reliably extract it.

\begin{figure}[b]
  \centering
  \includegraphics[width=0.7\hsize]{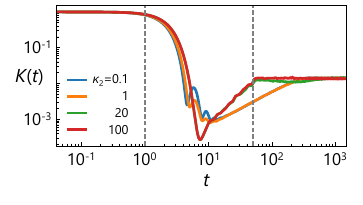}
  \caption{
    The SFF for the rescaled Hamiltonian \(\mathcal{H}\rightarrow\mathcal{H}/||\mathcal{H}||\) at several coupling strength \(\kappa_2\), with a fixed system size of \(N=8\) and \(J_4=1\). 
    The dashed lines signify \(t=1\) and \(50\), in relation to the input interval \(\Delta t_{\mathrm{in}}\) used in to evaluate QRC performance. 
    The results are averaged over \(20000\) realizations.
    }
  \label{figS4}
\end{figure}

When interpolating between the two models, the overall energy scale varies with the coupling strength. 
To compare QRC performance on an equal footing, we normalize the Hamiltonian by its spectral norm, \(\mathcal{H}\rightarrow\mathcal{H}/||\mathcal{H}||\). 
This normalization leaves the level spacing ratio invariant, as well the chaotic nature of the system. 
However, it introduces a uniform prefactor in the temporal domain, affecting the SFF [Eq.~(\ref{Seq3})] and the \(\Delta t_{\mathrm{in}}\) dependency of QRC performance. 
Figure \ref{figS4} displays the SFF for the interpolated model after the rescaling. 
In the quantum chaotic regime (\(\kappa_2=0.1\), \(1\)), the typical ramp-plateau profiles emerge, while in the integrable regime (\(\kappa_2=20\), \(100\)), the SFF dips and plateaus at much earlier times. 
Apart from the overall time-scale shift introduced by the Hamiltonian rescaling, these behaviors mirror those of the SYK\(_4\) and SYK\(_2\) models in Fig.~2(b) of the main text. 

In our QRC performance benchmarks in the parametric domain (Fig.~4 of the main text), we selected the input time intervals \(\Delta t_{\mathrm{in}}=50\) and \(\Delta t_{\mathrm{in}}=1\); the corresponding times are marked by the dashed lines in Fig.~\ref{figS4}. 
For the quantum chaotic regime, the longer \(\Delta t_{\mathrm{in}}\) falls within the RMT ramping window, while the shorter one precedes the dip. 
Similarly for the integrable models, these two times effectively sandwich the dip. 
These observations validate both the Hamiltonian normalization procedure and our choice of \(\Delta t_{\mathrm{in}}\) for probing distinct temporal regimes relevant to QRC. 
It is worth noting that precisely characterizing chaos in the interpolated model via the SFF typically requires a more sophisticated procedure called unfolding~\cite{Nosaka:JHEP:2018}. 
Unfolding recasts the energy spectrum such that it will exhibit a constant level spacing; it is not an operation directly applied to the Hamiltonian itself. 
We do not employ this procedure because our primary motivation is to characterize QRC performance by focusing on the properties of the original, physically implemented Hamiltonian.

\section{Dynamics of quantum reservoir system}

\begin{figure}[b]
  \centering
  \includegraphics[width=\hsize]{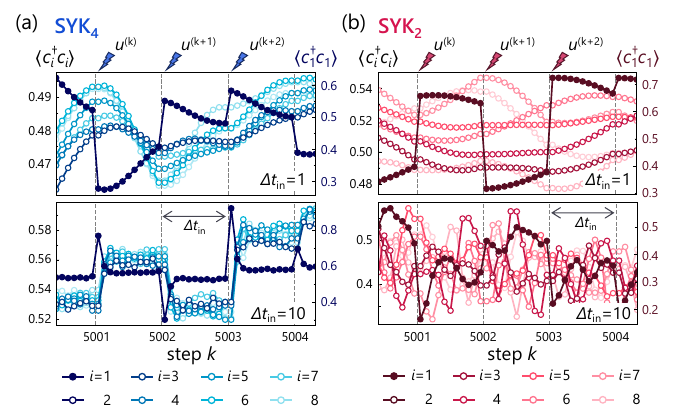}
  \caption{
    (a), (b) Dynamics of the read-out signals \(\langle c_i^\dagger c_i\rangle\) for the SYK\(_4\) model and the SYK\(_2\) model calculated for one typical realization. 
    The upper panels correspond to the input interval \(\Delta t_{\mathrm{in}}=1\), while the lower panels use \(\Delta t_{\mathrm{in}}=10\). 
    The vertical dashed lines indicate the input-encoding events \(\rho^{(k-1)}(\Delta t_{\mathrm{in}})\rightarrow \rho^{(k)}(0)\). 
    Data markers denote the values of \(\langle c_i^\dagger c_i\rangle\) at times \(v\Delta t_{\mathrm{in}} /V\) where \(0\le v\le V\). 
    The filled symbol corresponds to the value at the input site (scale on the right vertical axis), and the open symbols show values at the other sites (scale on the left vertical axis).
    }
  \label{figS5}
\end{figure}

The QRC protocol consists of signal encoding, reservoir dynamics, read-out, and post-processing (see Fig.~1 of the main text). 
We briefly overview this protocol here to provide deeper insight into its dynamics. 
First, for a sequential input \(\{u^{(k)}\}\), the \(k\)-th input \(u^{(k)}\) is encoded into the \(N\)-site quantum reservoir system by replacing the quantum state at site \(1\). 
The system then evolves unitarily under the Hamiltonian for a duration \(\Delta t_{\mathrm{in}}\) until the next input. 
During this evolution, the single-site particle number \(\langle c_i^\dagger c_i\rangle\) is measured at \(V\) subintervals. 
The \(NV\) resulting read-out signals, along with an additional constant term, form the \((NV+1)\)-dimensional state vector \(\bm{\mathrm{x}}^{(k)}\). 
The final output \(y^{(k)}\) is obtained through a linear transformation of this vector with the trainable weight \(W_{\mathrm{out}}\). 

Figures \ref{figS5}(a) and \ref{figS5}(b) illustrate the dynamics of read-out signals for the SYK\(_4\) and SYK\(_2\) models, respectively, for input intervals of \(\Delta t_{\mathrm{in}}=1\) and \(\Delta t_{\mathrm{in}}=10\). 
The dashed lines indicate when new inputs are provided. 
Markers show \(\langle c_i^\dagger c_i\rangle\) at times \(v\Delta t_{\mathrm{in}} /V\) (\(v=0,\dots,V\)), with the filled marker specifically denoting site \(1\), i.e., the input site. 
The dynamics at the input site change abruptly after an input, showing a larger variation range than other sites; hence, its vertical axis is displayed separately.

With a short input interval (\(\Delta t_{\mathrm{in}}=1\)), each new input arrives before the previous one has fully influenced the system. 
Thus, the resulting dynamics is relatively simple, as shown in the upper panels of both Figs.~\ref{figS5}(a) and \ref{figS5}(b). 
In contrast, system-dependent characteristics emerge when the input interval is longer (\(\Delta t_{\mathrm{in}}=10\)). 
In the SYK\(_4\) model [lower panel of Fig.~\ref{figS5}(a)], all readout signals tend toward quasi-stationary values, whereas in the SYK\(_2\) model [lower panel of Fig.~\ref{figS5}(b)], the signals exhibit continuous oscillations. 
Notably, although both dynamics appear sufficiently rich, their QRC performance differs significantly, as demonstrated in Figs.~3(c) and 3(d) of the main text. 
Because the output \(y^{(k)}\) is computed from these readout signals, this performance gap must originate from the difference in dynamics. 
However, the underlying reason for this performance gap is not immediately obvious from the raw signals in Fig.~\ref{figS5}.

These \(\Delta t_{\mathrm{in}}\)-dependent behaviors raise interest in the relationship between QRC performance and scrambling time---the timescale over which information delocalizes~\cite{Gharibyan:JHEP:2018}. 
Since scrambling relates to the propagation of encoded input information, its connection to memory performance is naturally anticipated~\cite{Kobayashi:SciPostPhys:2025}. 
Accordingly, the observed \(\Delta t_{\mathrm{in}}\) dependence of STM performance [upper panels of Figs.~3(c) and 3(d) of the main text] could plausibly reflect the underlying scrambling timescale. 
However, the pronounced enhancement in nonlinear processing performance [lower panel of Fig.~3(c) of the main text] cannot be explained by scrambling alone; scrambling itself is primarily associated with memory. 
Moreover, the convergence of QRC performance toward that of the Haar QRC model when \(\Delta t_{\mathrm{in}}\) exceeds the Thouless time underscores a fundamental connection to RMT. 
These observations collectively suggest that the Thouless time plays the more decisive role in determining overall QRC performance. 
Nevertheless, the detailed relationship between scrambling time, Thouless time, and the temporal edge of many-body quantum chaos warrants further investigation. 

\section{Convergence property}

\begin{figure}[b]
  \centering
  \includegraphics[width=\hsize]{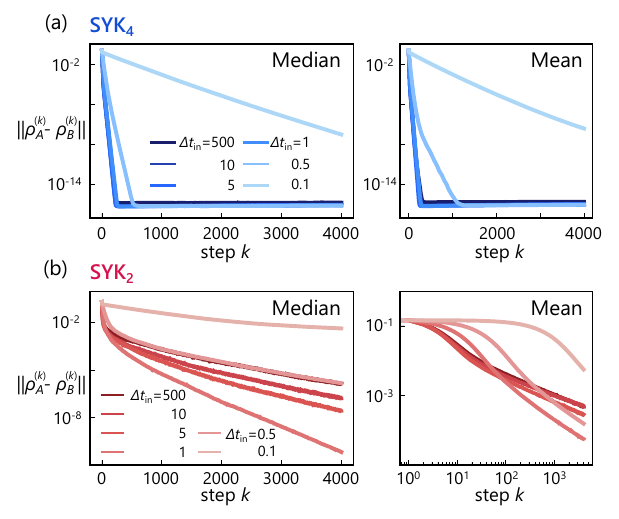}
  \caption{
    Evolution of the distances \(||\rho_A^{(k)}- \rho_B^{(k)}||\) as a function of the step \(k\) for (a) the SYK\(_4\) model and (b) the SYK\(_2\) model. 
    The left panels show the median distance (semilog scale), and the right panels show the mean distance [semilog for (a) and log-log for (b)], calculated from \(500\) independent pairs of initial states \(\rho_A^{(0)}\) and \(\rho_B^{(0)}\). 
    Colors indicate the input interval \(\Delta t_{\mathrm{in}}\). 
    }
  \label{figS6}
\end{figure}

We here examine echo state property (ESP), a fundamental characteristic of RC systems. 
ESP dictates that the reservoir state should asymptotically converge when driven by the same input sequence, irrespective of its initial conditions. 
In other words, after sufficiently many input injections, any dependence on the initial state should vanish. 
In QRC paradigms, ESP is typically assessed by measuring the distance between two quantum states, \(\rho_A^{(k)}\) and \(\rho_B^{(k)}\). 
These states evolve from different initial states, \(\rho_A^{(k=0)}\) and \(\rho_B^{(k=0)}\), but under an identical input sequence. 
ESP is considered satisfied when this distance approaches 0. 
Under purely unitary dynamics, the distance between these states, evaluated using the Frobenius norm, remains constant over time: \(||\rho_A^{(k)}(t) - \rho_B^{(k)}(t) || = ||\rho_A^{(k)}(t') - \rho_B^{(k)}(t') || \). 
Thus, the specific time \(t\) within a unitary evolution step is irrelevant. 
However, the distance does change upon the nonunitary input injection process, which involves the partial trace over site \(1\) and the reinitialization with the input state \(\rho_{\mathrm{in}}\). 
Consequently, to evaluate ESP, we monitor the state distance as a function of the input step \(k\). 
As the system traces out the previous state and injects a new, identical input state for both trajectories, the distance between \(\rho_A^{(k)}\) and \(\rho_B^{(k)}\) is expected to decrease with \(k\). 
Yet, the precise convergence behavior depends on the state before the input and the time evolution process \(e^{-i\mathcal{H}\Delta t_{\mathrm{in}}}\). 
For instance, if the input interval is close to zero (\(\Delta t_{\mathrm{in}} \rightarrow 0\)), the newly added input state is traced out almost immediately, and the distance remains largely unchanged. 
Conversely, a larger \(\Delta t_{\mathrm{in}}\) yields more intricate dynamics, where the partial trace operation can lead to a more significant reduction in distance, the details of which depend on the Hamiltonian. 

Figures \ref{figS6}(a) and \ref{figS6}(b) display the distance \(||\rho_A^{(k)} - \rho_B^{(k)} ||\) as a function of the input step \(k\) for various input intervals \(\Delta t_{\mathrm{in}}\) in the SYK\(_4\) and SYK\(_2\) models, respectively. 
The initial states \(\rho_{A,B}^{(0)}\) are randomly sampled from the half-filling sector. 
To accommodate values spanning several orders of magnitude, both the mean and median distances over \(500\) pairs of initial states are presented. 
For all models investigated, our results strongly indicate that the update rule [Eq.~(2) of the main text] is contractive for random initial conditions and inputs: \(||\rho_A^{(k+1)} - \rho_B^{(k+1)} || < ||\rho_A^{(k)} - \rho_B^{(k)} || \). 
This observation suggests that ESP is satisfied for these systems---at least asymptotically as \(k\rightarrow\infty\). 

The convergence speeds of the SYK\(_4\) and SYK\(_2\) models differ significantly. 
In the SYK\(_4\) model, the distance metric consistently decays exponentially toward \(0\), but the decay rate is sensitive to the input interval \(\Delta t_{\mathrm{in}}\). 
For smaller intervals (\(\Delta t_{\mathrm{in}}=0.1\), \(0.5\)), both the mean and median distances converge relatively slowly due to less effective erasure of past states upon the partial trace. 
Conversely, larger \(\Delta t_{\mathrm{in}}\) values result in much faster convergence, with the distance behaving similarly across these larger input intervals. 
In stark contrast, the SYK\(_2\) model exhibits considerably slower distance decay. 
While its median distance also decreases exponentially, the mean distance decays only polynomially and remains relatively high. 
This behavior reflects the integrable nature of the SYK\(_2\) model, wherein numerous internal conserved quantities can impede convergence. 
These conserved quantities also generate large statistical fluctuations, which explains the disparity between the mean and median distances [Fig.~\ref{figS6}(b)] and the large standard deviation in the performance metric (the SYK\(_2\)-dominated regime in Fig.~4 of the main text).

\begin{figure}[t]
  \centering
  \includegraphics[width=0.75\hsize]{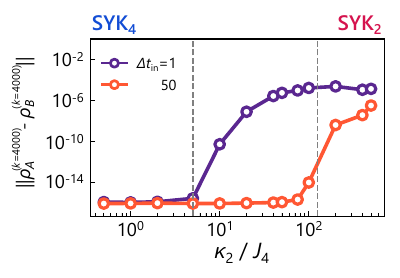}
  \caption{
    Median distance \(||\rho_A^{(k)}- \rho_B^{(k)}||\) at \(k=4000\) versus the coupling strength \(\kappa_2/J_4\). 
    Medians are calculated from \(500\) independent pairs of random initial states \(\rho_A^{(0)}\) and \(\rho_B^{(0)}\), and are presented for two input intervals: \(\Delta t_{\mathrm{in}}=1\) (purple) and \(\Delta t_{\mathrm{in}}=50\) (orange). 
    The vertical lines indicate the onset of WD (left) and Poisson (right) statistics. 
    }
  \label{figS7}
\end{figure}

\begin{figure*}[!t]
  \centering
  \includegraphics[width=0.95\hsize]{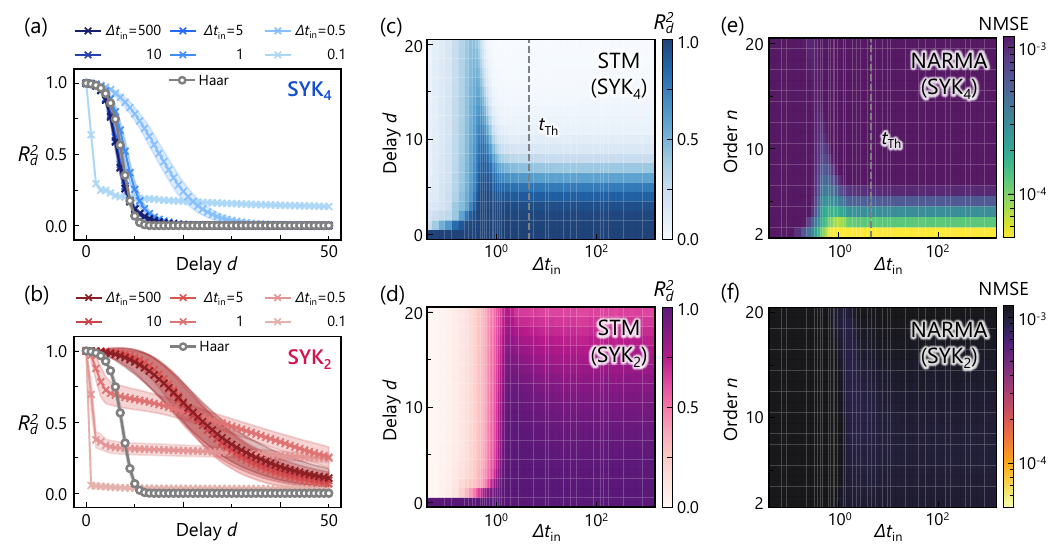}
  \caption{
    (a), (b) QRC performance for the STM task \(R^2_d\) as a function of delay \(d\) for various input intervals \(\Delta t_{\mathrm{in}}\), with shaded regions indicating standard deviations. 
    The gray curve represents the Haar QRC performance. 
    (c), (d) Heatmaps illustrating STM task performance, where the vertical axis represents the delay step \(d\) and the horizontal axis corresponds to the input interval \(\Delta t_{\mathrm{in}}\).
    (e), (f) Heatmaps of NARMA task performance. 
    The vertical axis denotes the NARMA order \(n\), and the horizontal axis shows the input interval \(\Delta t_{\mathrm{in}}\). 
    Results for the SYK\(_4\) model are presented in (a), (c), (e), while results for the SYK\(_2\) model are in (b), (d), (f). 
    The Thouless time \(t_{\mathrm{Th}}\) is marked by vertical dashed lines in (c) and (e).
    }
  \label{figS8}
\end{figure*}

Figure \ref{figS7} displays the median distance \(||\rho_A^{(k)} - \rho_B^{(k)} ||\) for the interpolated model as a function of the coupling strength \(\kappa_2/J_4\). 
The distance is evaluated at the final step considered (\(k=4000\)) and is presented for two input intervals: \(\Delta t_{\mathrm{in}}=1\) and \(\Delta t_{\mathrm{in}}=50\). 
Larger final distances indicate slower convergence, whereas smaller ones suggest rapid decay. 
Consistent with Figs.~\ref{figS6}(a) and \ref{figS6}(b), the distance approaches \(0\) for small \(\kappa_2/J_4\) and remains relatively large in the strong \(\kappa_2/J_4\) regime for both input intervals. 
However, the behavior differs between the two input intervals at intermediate \(\kappa_2/J_4\) values. 
For \(\Delta t_{\mathrm{in}}=1\), the distance deviates from \(0\) at relatively small \(\kappa_2/J_4\), near the onset of WD statistics. 
In contrast, with \(\Delta t_{\mathrm{in}}=50\), it stays close to \(0\) until \(\kappa_2/J_4\) approaches the boundary of the Poisson regime. 
This comparison indicates that convergence becomes harder to achieve as the system nears the integrable regime, which requires a large \(\Delta t_{\mathrm{in}}\) to generate the necessary complex dynamics. 

We note that Ref.~\cite{Pena:PRL:2021} analyzes QRC performance in spin-based quantum reservoirs, attributing the poor performance observed in localized phases to slow convergence. 
Our QRC likewise exhibits slow convergence in the SYK\(_2\)-dominant regime (Fig.~\ref{figS7}). 
Nevertheless, as shown in Fig.~4(a) of the main text, the parametric edge of many-body quantum chaos appears at \(\kappa_2/J_4\) values almost an order of magnitude smaller than those at which the distance metric first deviates from \(0\) in Fig.~\ref{figS7}. 
Moreover, although the convergence metric varies significantly with  \(\kappa_2/J_4\) even for \(\Delta t_{\mathrm{in}}=1\), the performance peak associated with the parametric edge does not appear in Fig.~4(b) of the main text. 
Taken together, these observations suggest that convergence and the parametric edge are not directly related here. 

\section{Reservoir performance}

The main text presents QRC performance on the short-term memory (STM) task and the nonlinear auto-regressive moving average (NARMA) task. 
Specifically, it benchmarks the SYK\(_4\) [Fig.~3(c)] and SYK\(_2\) [Fig.~3(d)] models in the temporal domain, as well as the interpolated model in the parametric domain at input intervals of \(\Delta t_{\mathrm{in}}=50\) [Fig.~4(a)] and \(\Delta t_{\mathrm{in}}=1\) [Fig.~4(b)]. 
The aim here is to provide additional data to further elucidate QRC performance and the edge of many-body quantum chaos. 
In Secs.~\ref{sec:S4-1} and \ref{sec:S4-2}, we examine the STM and NARMA performance in the temporal and parametric domains, respectively, by sweeping the delay parameters and task orders. 
We also report the QRC performance on additional tasks: the nonlinear STM task in Sec.~\ref{sec:S4-3} and the parity-check (PC) task in Sec.~\ref{sec:S4-4}. 
Finally, Sec.~\ref{sec:S4-5} investigates the effects of statistical noise on QRC performance. 

\subsection{Reservoir performance for the SYK\(_4\) and SYK\(_2\) models}\label{sec:S4-1}

This section examines the detailed QRC performance of the SYK\(_4\) and SYK\(_2\) models in the temporal domain. 
We first analyze performance on the STM task, which quantifies linear memory capability. 
Figures~\ref{figS8}(a) and \ref{figS8}(b) present the memory performance \(R^2_d\) as a function of delay \(d\) for each model, while the input interval \(\Delta t_{\mathrm{in}}\) is varied. 
The performance of the Haar QRC model is included for baseline comparison. 
For the SYK\(_4\) model, the performance \(R^2_d\) approaches the Haar QRC reference value at large \(\Delta t_{\mathrm{in}}\), reflecting a universal RMT description when \(\Delta t_{\mathrm{in}}\) exceeds the Thouless time \(t_{\mathrm{Th}}\). 
Optimal performance in this plot occurs at \(\Delta t_{\mathrm{in}}=0.5\), before the Thouless time, where a long tail of nonzero \(R^2_d\) extends to \(d=30\). 
Conversely, a very short input interval (\(\Delta t_{\mathrm{in}}=0.1\)) yields poor performance, with \(R^2_d\) remaining around \(0.2\) even at large \(d\). 
This degradation arises because the partial trace provides less effective memory erasure for such short input intervals, leading to the slow convergence observed in Fig.~\ref{figS7}(a). 

The tendency of short input intervals to introduce atypical decay in \(R^2_d\) is more pronounced in the SYK\(_2\) model [Fig.~\ref{figS8}(b)]. 
For instance, at \(\Delta t_{\mathrm{in}}\leq 1\), the memory performance drops abruptly instead of decaying gradually. 
This behavior suggests a compromised fading-memory property; due to the slow convergence, the system fails to sufficiently prioritize recent information. 
Nevertheless, for longer input intervals (\(\Delta t_{\mathrm{in}}=5\), \(10\), and \(500\)), QRC with the SYK\(_2\) model demonstrates significantly better memory retention than the Haar QRC baseline, with a long tail spanning beyond \(d=50\). 
Importantly, this enhancement appears even though the convergence speed [quantified by \(||\rho_A^{(k)}- \rho_B^{(k)}||\) in Fig.~\ref{figS6}(b)] is comparable to that observed at shorter \(\Delta t_{\mathrm{in}}\). 
This discrepancy implies that, while the distance-based convergence metric is valuable for characterizing QRC systems, its direct connection to performance can be system-dependent. 
Specifically, in the SYK\(_2\) model, ESP is satisfied only asymptotically, and it remains nontrivial to determine how this asymptotic nature precisely affects performance. 
Further investigation, potentially employing alternative convergence metrics, is left for future work. 

Figures \ref{figS8}(c) and \ref{figS8}(d) extend Figs.~\ref{figS8}(a) and \ref{figS8}(b), respectively, by presenting heatmaps of \(R_d^2\) for \(0\leq d\leq 20\) as the input interval \(\Delta t_{\mathrm{in}}\) is varied. 
In the SYK\(_4\) model [Fig.~\ref{figS8}(c)], a key observation is an oblique line of high performance that appears where \(\Delta t_{\mathrm{in}}\) is shorter than the Thouless time. 
This sharp feature represents the performance enhancement at the temporal edge of many-body quantum chaos, consistent with the observation in Fig.~3(c) of the main text. 
Notably, as discussed there, the exact location of the performance peak is task-dependent. 
For the STM task investigated here, the peak shifts gradually to the nonchaotic side (smaller \(\Delta t_{\mathrm{in}}\)) as \(d\) increases, leading to the observed oblique line. 
By contrast, a similar enhancement is absent in the SYK\(_2\) model [Fig.~\ref{figS8}(d)]; instead, it shows a binary separation between a regime of sufficient memory retention at large \(\Delta t_{\mathrm{in}}\) and poor retention at small \(\Delta t_{\mathrm{in}}\), lacking any distinct performance peak. 

Figures~\ref{figS8}(e) and \ref{figS8}(f) show heatmaps of the NMSE for the NARMA tasks using the SYK\(_4\) and SYK\(_2\) models, respectively. 
The vertical axis shows the NARMA order \(n\), and the horizontal axis indicates the input interval \(\Delta t_{\mathrm{in}}\). 
Consistent with the STM task results, a distinct performance peak is evident in the SYK\(_4\) model, whereas the SYK\(_2\) model shows only gradual performance changes. 
As discussed in the main text, the absence of performance peak in the SYK\(_2\) model indicates that the performance enhancement observed in the SYK\(_4\) model indeed originates from many-body quantum chaos. 
Notably, as the NARMA order \(n\) increases, the peak broadens and eventually vanishes, illustrating a regime in which the edge-induced peak vanishes due to excessive task complexity.

\begin{figure*}[!t]
  \centering
  \includegraphics[width=0.95\hsize]{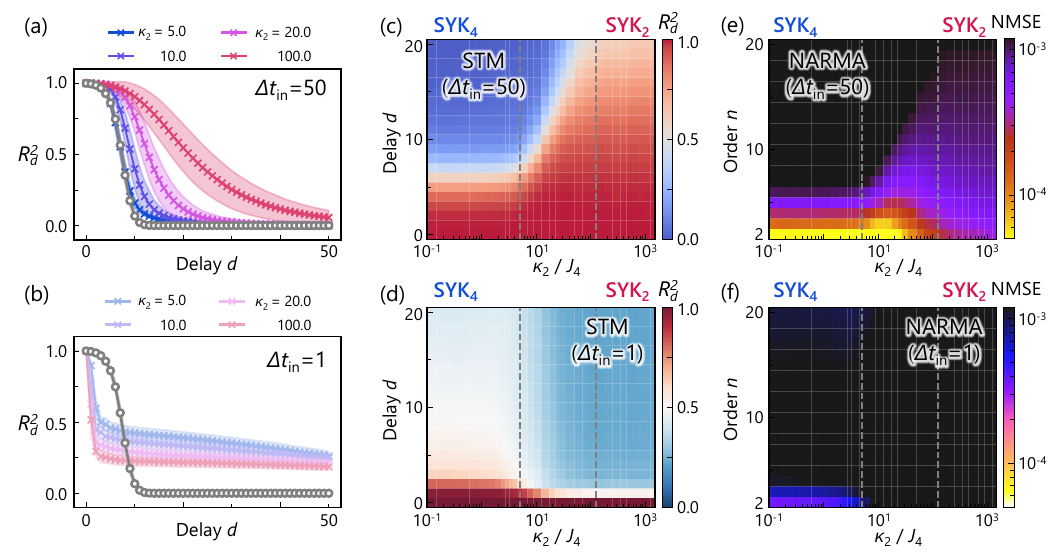}
  \caption{
    (a), (b) QRC performance on the STM task \(R^2_d\) as a function of delay \(d\). 
    Different colored lines represent various coupling strengths \(\kappa_2\), with shaded areas denoting standard deviations. 
    The gray curve represents the Haar QRC performance. 
    (c), (d) Heatmap representation of the STM task performance \(R^2_d\). 
    The vertical axis represents the delay step \(d\), and the horizontal axis corresponds to the ratio of coupling strength \(\kappa_2/J_4\).
    (e), (f) Heatmaps of NARMA task performance. 
    The vertical axis denotes the NARMA order \(n\), while the horizontal axis shows the coupling strength \(\kappa_2/J_4\). 
    In heatmaps (c)-(f), vertical dashed lines mark the approximate onset of WD (left) and Poisson (right) statistical distributions in the spectral properties.
    Panels (a), (c), (e) correspond to the input interval of \(\Delta t_{\mathrm{in}}=50\), while panels (b), (d), (f) correspond to \(\Delta t_{\mathrm{in}}=1\). 
    }
  \label{figS9}
\end{figure*}

\subsection{Reservoir performance for the interpolated model}\label{sec:S4-2}

Next, we analyze QRC performance of the model interpolating between the SYK\(_4\) and SYK\(_2\) systems in Fig.~\ref{figS9}, extending the results presented in Fig.~4 of the main text. 

Figures~\ref{figS9}(a) and \ref{figS9}(b) illustrate the memory performance \(R^2_d\) as a function of delay \(d\) for various coupling strengths \(\kappa_2\) at \(J_4=1\). 
For comparative purposes, the performance of the Haar QRC model is also plotted. 
In the long time-scale regime (\(\Delta t_{\mathrm{in}}=50\)), the QRC performance within the quantum chaotic regime (\(\kappa_2=5.0\)) aligns with the Haar QRC reference value, which is a direct consequence of universal RMT behavior [Fig.~\ref{figS9}(a)]. 
As the coupling strength \(\kappa_2\) increases, the STM performance improves monotonically, eventually exhibiting an extended tail in \(R^2_d\) in the integrable regime (\(\kappa_2=100.0\)). 
This trend is further elucidated in the corresponding heatmap in Fig.~\ref{figS9}(c), which plots \(R^2_d\) against the coupling strength ratio \(\kappa_2 / J_4\) (horizontal axis) and delay \(d\) (vertical axis). 
The onsets of the chaotic and integrable regimes are indicated by gray dashed lines. 
Within the quantum chaotic regime, the performance is nearly uniform across \(\kappa_2 / J_4\), showing a similar dependence on delay \(d\). 
Upon entering the intermediate regime, \(R^2_d\) gradually improves towards the integrable regime, where memory retention is strongest. 
The boundary separating high (red) and low (blue) \(R^2_d\) values forms an oblique line, indicating that stronger \(\kappa_2\) coupling is necessary to preserve memory for longer delays \(d\). 

By contrast, in the short time-scale regime (\(\Delta t_{\mathrm{in}}=1\)), \(R^2_d\) displays a markedly different pattern: a sharp initial drop followed by a slow decay, the decay rate of which diminishes further as the system approaches the integrable limit at large \(\kappa_2\) [Fig.~\ref{figS9}(b)]. 
The heatmap in Fig.~\ref{figS9}(d) corroborates this observation: a region of high \(R^2_d\) (red) is concentrated at small delays (bottom), illustrating the initial high performance before an abrupt drop. 
The subsequent slow decay corresponds to the upper regions of nearly uniform, low \(R^2_d\) (blue). 
These behaviors are attributed to the less effective memory erasure via the partial trace operation when the input interval \(\Delta t_{\mathrm{in}}\) is short, as discussed previously.

We additionally present performance heatmaps for the NARMA task in Fig.~\ref{figS9}(e) for \(\Delta t_{\mathrm{in}}=50\) and Fig.~\ref{figS9}(f) for \(\Delta t_{\mathrm{in}}=1\). 
The horizontal axis represents the coupling strength \(\kappa_2 / J_4\), and the vertical axis shows the NARMA order \(n\). 
For small NARMA orders with a long input interval [Fig.~\ref{figS9}(e)], the performance is distinctly enhanced around the chaotic boundary, corresponding to the parametric edge of many-body quantum chaos. 
This edge-induced performance enhancement is absent for the short input interval \(\Delta t_{\mathrm{in}}=1\) in Fig.~\ref{figS9}(f) because, as discussed in the main text, the QRC requires a sufficiently long input interval to manifest its quantum chaotic characteristics. 
As the NARMA order \(n\) increases, the task becomes substantially more challenging, and the QRC can no longer represent it adequately, as evidenced by the degraded performance. 
In this regime with large \(n\), the effective strategy is essentially limited to prioritizing the representation of contributions driven by distant past inputs, rather than capturing complex nonlinear details. 
Accordingly, in the high-\(n\) regime the performance contours for the NARMA task [Fig.~\ref{figS9}(e)] resemble those for the STM task [Fig.~\ref{figS9}(c)].

\subsection{Nonlinear STM task}\label{sec:S4-3}

To further benchmark the QRC, we employ the nonlinear extension of the STM task~\cite{Llodra:AdvQuantTech:2023,Llodra:Chaos:2025}. 
This task is designed to systematically vary the relative demands on memory and nonlinear processing. 
As in the standard STM task, a sequence of random inputs \(\{u^{(k)}\}\), with \(u^{(k)}\in[0,1]\), is provided to the quantum reservoir. 
For a given polynomial order \(n\) and delay \(d\), the target output at time step \(k\) is defined as \(\bar{y}^{(k)} = \left(u^{(k-d)}\right)^n\). 
The case \(n=1\) recovers the standard (linear) STM task analyzed in the main text. 
By increasing \(n\) at fixed \(d\), the task selectively enhances the demand for nonlinear processing, while the memory requirement (determined by \(d\)) remains unchanged.
The QRC performance on this task is quantified using the determination coefficient \(R^2_d\).

We present the performance on the nonlinear STM task with delay \(d=4\) around the temporal and parametric edges of many-body quantum chaos in Figs.~\ref{figS10}(a) and \ref{figS10}(b), respectively. 
In particular, the temporal edge is studied using the SYK\(_4\) model, while the parametric edge is examined using the interpolated model at \(\Delta t_{\mathrm{in}}=50\). 
In both cases, the performance decreases monotonically with increasing the nonlinearity order \(n\), reflecting the greater task complexity. 
Notably, in the temporal domain, the distinct performance peak at the temporal edge persists from low- to high-order nonlinearities. 
The position of this peak appears to be independent of the nonlinearity order \(n\), suggesting that the location of the temporal edge for this task is governed by the memory properties.

In the parametric domain, the performance peak is blurred for the linear STM task (\(n=1\)), because memory retention is strongest in the integrable regime. 
However, as the nonlinearity is introduced to the task via \(n\geq2\), QRC performance in the integrable regime degrades more rapidly than in the chaotic regime, reflecting the former's limited nonlinear transformation capability. 
Consequently, a performance peak at the parametric edge of many-body quantum chaos becomes prominent for these tasks. 
Remarkably, with increasing \(n\), the peak position shift towards smaller \(\kappa_2/J_4\) values, i.e., closer to the chaotic regime. 
This shift reflects that operating closer to the chaotic side enhances the nonlinear transformation capacity, which is required to solve tasks with higher nonlinearity order \(n\).

\begin{figure}[tbhp]
  \centering
  \includegraphics[width=\hsize]{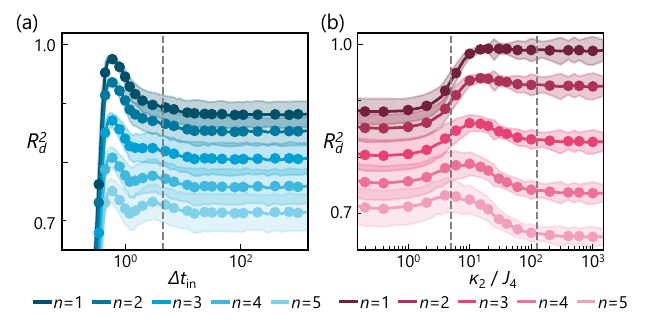}
  \caption{
    QRC performance on the nonlinear STM task with delay \(d=4\) for orders \(n=1\) to \(5\). 
    (a) Performance of the SYK\(_4\) model plotted as a function of the input interval \(\Delta t_{\mathrm{in}}\). 
    The Thouless time \(t_{\mathrm{th}}\) is indicated by the vertical dashed line. 
    (b) Performance of the interpolated model as a function of the coupling strength \(\kappa_2/J_4\), with the input interval fixed at \(\Delta t_{\mathrm{in}}=50\). 
    The onsets of the chaotic and integrable regimes are marked by the left and right horizontal lines, respectively.
    }
  \label{figS10}
\end{figure}

\subsection{PC task}\label{sec:S4-4}

We next evaluate the PC task~\cite{Fujii:PRAppl:2017, Xia:FrontPhys:2022, Sannia:Quantum:2024, Llodra:Chaos:2025}. 
In this task, binary inputs \({u'}^{(k)}\in\{0,1\}\) are injected into the quantum reservoir, and the target output is the parity of the sum of the previous inputs: \(\bar y^{(k)}=\left(\sum_{d=0}^{n} u'^{(k-d)}\right)\bmod 2\). 
We also use the determination coefficient \(R^2\) to evaluate performance on this task. 
A defining feature of this task is its strong nonlinearity requirement. 
The PC task of order \(n\) requires memory spanning \(n+1\) steps (the current input plus the \(n\) preceding ones). 
More importantly, it requires nonlinearity of order \(n+1\) involving many cross-term interactions among delayed inputs, because the parity function corresponds to an (\(n+1\))-degree polynomial in its inputs. 
In other words, increasing \(n\) by one adds only a single additional step to the memory requirement, whereas the nonlinearity requirement grows much more sharply both in polynomial order and in the combinatorial complexity of cross-term interactions.

\begin{figure}[b!]
  \centering
  \includegraphics[width=\hsize]{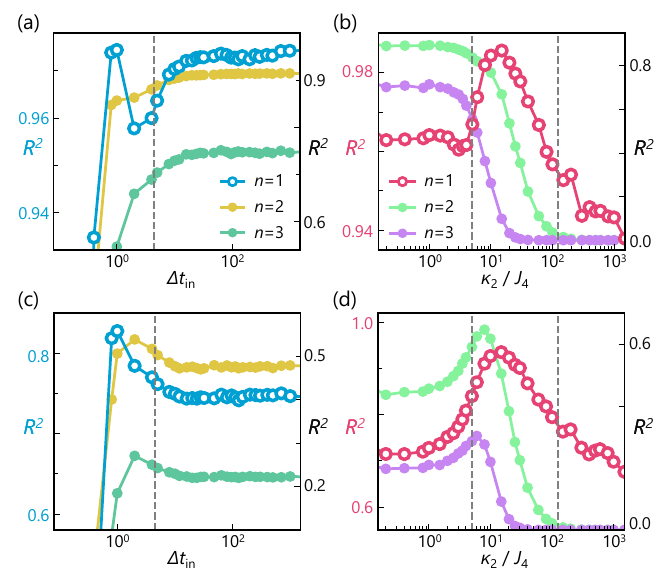}
  \caption{
    QRC performance on (a), (b) the PC task and (c), (d) the PC task with memory bias \(b=2\). 
    (a), (c) Performance using the SYK\(_4\) model as a function of \(\Delta t_{\mathrm{in}}\). 
    (b), (d) Performance using the interpolated model as a function of \(\kappa_2/J_4\), with \(\Delta t_{\mathrm{in}} = 50\). 
    In all panels, \(R^2\) values for \(n=1\) are plotted on the left vertical axis, and values for \(n=2\) and \(3\) are on the right vertical axis. 
    }
  \label{figS11}
\end{figure}

Figures \ref{figS11}(a) and \ref{figS11}(b) presents the QRC performance for the PC task with \(n=1,2\), and \(3\), evaluated in the temporal domain [Fig.~\ref{figS11}(a)] and the parametric domain [Fig.~\ref{figS11}(b)]. 
For \(n=1\), we observe clear performance peaks at the edge of many-body quantum chaos in both cases. 
In contrast, for \(n=2\) and \(3\), the task's nonlinearity requirement escalates sharply and overwhelms the memory demand. 
Consequently, optimal performance is achieved within the chaotic regime, where the nonlinearity of the input-to-output mapping is strongest. 
This behavior exemplifies an extreme imbalance in task requirements: when either memory or nonlinear transformation dominates, the optimal operating regime is integrable or chaotic, respectively, and the edge-induced performance peak is absent. 
The linear STM task (memory-dominated) and the PC task with \(n\ge 2\) (nonlinearity-dominated) exemplify these extremes. 
It should be noted that in such limits, explicit design guidelines are less necessary because the optimal operating regime follows directly from the task definition.

Having observed that the edge-induced peak appears when a nonlinearity requirement is added to an otherwise memory-dominant STM task (the nonlinear STM task in Fig.~\ref{figS10}), we next examine whether introducing an additional memory requirement to the PC task likewise produces such a peak.
To this end, we modify the target output to \(\bar y^{(k)}=\left(\sum_{d=0}^{n} u'^{(k-d-b)}\right)\bmod 2\), where \(b\) denotes the temporal bias step. 
Figures \ref{figS11}(c) and \ref{figS11}(d) present the QRC performance for the PC task \(n=1,2, 3\) under the bias \(b=2\). 
Strikingly, a clear performance enhancement at the edge of many-body quantum chaos is observed for both edges, across all orders investigated. 
Because the bias  \(b\) increases only the memory requirement while leaving the nonlinearity unchanged, these results support the interpretation that the absence of a performance peak in the original PC task is due to the dominance of the nonlinear requirement. 
Introducing the bias makes memory a performance limiting factor alongside nonlinearity, thereby resolving the imbalance in task demands. 
The resulting setting thus becomes analogous to the NARMA task and the nonlinear STM task, for which both memory and nonlinearity constrain performance; in such nontrivial regimes, the edge of many-body quantum chaos becomes most prominent. 
Although not known a priori, realistic scenarios involving real-world data are expected to fall into this category. 
The pronounced performance peaks observed for these tasks therefore highlight the importance of operating at the edge of many-body quantum chaos as a robust design guideline for QRC. 
Finally, we note that increasing the order \(n\) in the biased PC task shifts the peak toward the chaotic regime, paralleling the trend observed in the nonlinear STM task (Fig.~\ref{figS10}).

\subsection{Impact of statistical
uncertainty in measurements}\label{sec:S4-5}

\begin{figure}[t!]
  \centering
  \includegraphics[width=\hsize]{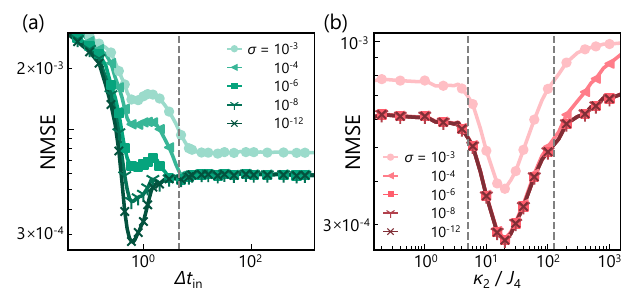}
  \caption{
    The QRC performance (NMSE) on the NARMA-5 task under sampling noise. 
    (a) The NMSE for the SYK\(_4\) model as a function of the input interval \(\Delta t_{\mathrm{in}} \). 
    (b) The NMSE for the interpolated model as a function of the coupling strength \(\kappa_2/J_4\), with \(\Delta t_{\mathrm{in}}\) fixed at \(50\). 
    In both panels, the colors and markers indicate the sampling noise strength \(\sigma\).
    }
  \label{figS12}
\end{figure}

Because expectation values are estimated from a finite number of measurement shots, their accuracy is limited by sampling noise. 
Our analysis thus far has assumed ideal, noise-free conditions; we now investigate the impact of statistical noise on QRC performance by modeling finite sampling error as additive Gaussian noise. 
Specifically, to the noiseless expectation value \(n_i=\langle c_i^\dagger c_i\rangle\), we add a zero-mean Gaussian noise with standard deviation \(\sigma_i=\sigma\sqrt{n_i(1-n_i)}\), where \(\sigma\) parametrizes the shot noise strength~\cite{Mujal:npjqi:2023,Palacios:CommPhys:2024,Kobayashi:PRXQuantum:2024}.  
This variance matches the binomial variance of repeated projective measurements in the \(c_i^\dagger c_i\) basis, and the Gaussian approximation is valid when \(\sigma\ll 1\).

In Figs.~\ref{figS12}(a) and \ref{figS12}(b), we show the NARMA-5 performance around the temporal edge and parametric edge, respectively, as the sampling noise strength \(\sigma\) is varied. 
As expected for algorithms that rely on estimating expectation values, statistical fluctuations degrade the performance; accordingly, reducing \(\sigma\) monotonically improves it. 
Note that the sensitivity to this noise depends on the operating regime. 
In the temporal sweep [Fig.~\ref{figS12}(a)], small input intervals \(\Delta t_{\mathrm{in}}\) are markedly more vulnerable than large \(\Delta t_{\mathrm{in}}\) (the RMT regime), because the system-specific structure is more pronounced at small \(\Delta t_{\mathrm{in}}\) and is therefore more susceptible to statistical noise. 
For large \(\sigma\), this degradation overwhelms the edge-induced enhancement, producing local minima near the temporal edge and shifting the global minimum to the large-\(\Delta t_{\mathrm{in}}\) regime. 
As \(\sigma\) is reduced to \(10^{-8}\), the temporal edge again yields the global minimum; at \(\sigma=10^{-12}\), it nearly recovers the ideal performance shown in Fig.~3(c) of the main text. 
By contrast, in the parametric sweep we fix \(\Delta t_{\mathrm{in}}=50\), where the QRC is comparatively robust to sampling noise. 
Consequently, the parametric edge of many-body quantum chaos attains the global minimum even at \(\sigma=10^{-3}\), and the performance is essentially converged to the ideal from \(\sigma=10^{-4}\) onward.

\end{NoHyper}


\end{document}